\documentclass[11pt,a4paper]{article}
\usepackage[margin=0.95in]{geometry}
\usepackage{amsmath,amssymb,amsthm}
\usepackage{booktabs}
\usepackage{caption}
\usepackage{hyperref}
\usepackage{xcolor}
\usepackage{microtype}
\usepackage{tikz}
\usetikzlibrary{arrows.meta,positioning,shapes.geometric,calc}
\definecolor{linkblue}{HTML}{1E4E79}
\definecolor{accentgreen}{HTML}{136F4A}
\definecolor{accentamber}{HTML}{9A5B00}
\definecolor{accentred}{HTML}{9E2A2B}
\definecolor{mutedgray}{HTML}{4A5568}
\definecolor{boxbg}{HTML}{F7FAFC}
\definecolor{greenbg}{HTML}{EDF7F2}

\hypersetup{
  colorlinks=true,
  linkcolor=linkblue,
  citecolor=linkblue,
  urlcolor=linkblue,
  pdftitle={Packing Stern into L2: GPU-Accelerated Primal-Shortened ISD for High-Codimension Binary Goppa Key Recovery},
  pdfauthor={Raullen Chai, Xinxin Fan}
}

\title{\textbf{Packing Stern into L2:}\\
GPU-Accelerated Primal-Shortened ISD for High-Codimension Binary Goppa Key Recovery}

\author{Raullen Chai \qquad Xinxin Fan\\[3pt]
\small MachineFi Labs\\
\small \texttt{\{raullen, xinxin\}@machinefi.com}}
\date{November 23, 2026}

\begin{document}
\maketitle

\begin{abstract}
Primal-Shortened Information-Set Decoding (ISD) with quadratic polar lifting recently pushed TII McEliece Track-2 key recovery to $mt=126$ (\texttt{TII-228}, $2{,}463$ CPU core-hours), with over $99\%$ of runtime spent harvesting weight-$(2t+1)$ codewords. Porting MiniMax's Mode-6 Stern sampler directly to one block per SM leaves $1.0\text{--}1.5\text{ GB}$ of hash tables active and stalls on global memory. We pack $128$ syndrome bits into one $16$-byte \texttt{ulonglong2} without left-triple indices (lossless for $mt \le 145$, and extended to $mt \le 192$ via truncated range pre-filtering), share $G \in \{4, 8\}$ cooperative tables inside GPU L2 cache, batch $64G$ random systematic reductions per launch, and widen the shortening prefix $n'$ to cut modeled GPU-hours by $34\%\text{--}47\%$.

Across five Track-2 challenges ($mt = 108 \to 153$), every recovered private key passes both the official SageMath checker and a standalone pure-Python verifier. Eight Modal L40S GPUs solve \texttt{TII-228} ($mt=126$) in $18.9\text{ min}$ ($2.15$ GPU-hours, $3.0\times$ faster in wall-clock than MiniMax's EPYC cluster). We further recover keys for the previously undocumented challenges \texttt{TII-237} ($mt=135$, $22.1$ L4 GPU-hours), \texttt{TII-235} ($mt=144$, $39.62$ L40S GPU-hours, $17.97\text{ s}$ lift), and \texttt{TII-185} ($mt=153$, $140.4$ L40S + $\approx 60$ L4 GPU-hours, $39.01\text{ s}$ lift). These results do not demonstrate practical recovery against Classic McEliece parameters ($mt \ge 768$).

\end{abstract}

\section{Introduction}

The McEliece cryptosystem~\cite{McEliece78} conceals a binary Goppa code $\Gamma(L, g)$ behind a systematic parity-check matrix $H_{\mathrm{pub}} \in \mathbb{F}_2^{mt \times n}$. Although Classic McEliece reached Round~4 of the NIST post-quantum standardization process before NIST selected HQC in March 2025, concrete structural key-recovery margins on smaller parameters remained sparsely mapped until the Technology Innovation Institute (TII) published $204$ non-toy Track-2 public-key recovery challenges alongside two toy verification instances~\cite{TIIRepo}.

In November 2026, MiniMax~\cite{MiniMax26} extended Track-2 recoveries to $94$ keys up to codimension $mt=126$ using Primal-Shortened ISD (Route~A) alongside an auxiliary HOVER pipeline (Route~B). What makes Route~A structurally unusual is its runtime asymmetry. On \texttt{TII-245} ($mt=108$), SageMath recovers $(g, L)$ from harvested codewords in under ten seconds, whereas an OpenMP C sampler (\texttt{minwords\_mt.c}) consumes over $99\%$ of wall-clock time searching for weight-$(2t+1)$ vectors in a shortened parity-check kernel. The algebra is already cheap. Everything that limits codimension scaling sits inside a memory-bound collision search.

We leave MiniMax's algebraic recovery algorithm and Mode-6 Stern parameters ($p=3+3, l=17$, $16\text{ KB}$ occupancy bitmap) intact, and redesign the harvesting stage around two coupled constraints: GPU L2 cache capacity and the shortening-window trade-off. Sizing $G=4$ or $G=8$ cooperative hash tables below nominal L2 capacity reduces global-memory pressure; widening the shortened prefix $n'$ beyond narrow CPU-oriented choices trades a modest increase in $\mathbb{F}_2$ matrix dimension for a $10\text{--}25\times$ jump in codeword supply, yielding a modeled $34\%\text{--}47\%$ reduction in GPU-hours. Together, these changes compress \texttt{TII-228} ($mt=126$) from $57.1\text{ min}$ on $2{,}048\text{--}5{,}120$ EPYC cores to $18.9\text{ min}$ on eight L40S GPUs, and push verified Track-2 key recovery through \texttt{TII-237} ($mt=135$) and \texttt{TII-235} ($mt=144$, punctured at $10$ positions) to \texttt{TII-185} ($mt=153$).

Because codimension $mt$ does not order difficulty monotonically across arbitrary $(n, m, t, n')$ tuples---MiniMax's \texttt{TII-250} ($mt=120, k'=347$) required far more search work than \texttt{TII-237} ($mt=135, k'=243$)---we treat $mt=153$ as a concrete parameter milestone rather than a universal difficulty claim. In packing terms, \texttt{TII-235} ($mt=144$) occupies $127$ of $128$ bits in our lossless $16$-byte \texttt{ulonglong2} format (\texttt{v4}, ceiling $mt=145$), while \texttt{TII-185} ($mt=153$) crosses that ceiling by $8$ bits and validates our truncated $16$-byte range pre-filter (\texttt{v5}) without inflating the L2-sized hash table.

\section{Prior Work and the Track-2 Landscape}

During 2026, two distinct families of structural attacks drove Track-2 progress: pure algebraic vanishing methods and hybrid combinatorial-algebraic harvesting.

Bardet, Lemoine, and Tillich~\cite{BLT26} broke instances up to \texttt{TII-210} ($mt=63$) via Pfaffian relations, and Hemmert~\cite{Hemmert26} reached \texttt{TII-248} ($mt=63$) using higher-order vanishing (HOV). Following hold-out improvements by Vedenev~\cite{Vedenev26} and Weis~\cite{Weis26} on \texttt{TII-253} ($n=214, m=8, t=9, mt=72$), Saarinen introduced Frobenius-reduced HOVER to solve \texttt{TII-252} ($n=1{,}008, m=10, t=11, mt=110$)~\cite{Saarinen26HOVER} and Two-Anchor Holdout/Hermite to solve \texttt{TII-254} ($mt=96$, reporting $27.2$ GPU-hours for the two final Krylov sequences on $8\times$ NVIDIA GH200)~\cite{Saarinen26TwoAnchor}. Algebraic vanishing attacks require almost no combinatorial search, yet their matrix dimensions grow steeply with Goppa degree $t$ and code rate, leaving high-$t$ instances such as \texttt{TII-228} ($t=14$) through \texttt{TII-235} ($t=16$) beyond current HOVER systems.

\begin{figure}[t]
\centering
\begin{tikzpicture}[
  x=1.02cm, y=0.90cm,
  >=Stealth,
  node font=\scriptsize,
  pill/.style={draw, rounded corners=2pt, inner sep=2pt, font=\scriptsize\bfseries}
]
  \draw[->, mutedgray, thick] (0.2, 0.3) -- (13.6, 0.3);
  \draw[->, mutedgray, thick] (0.2, 0.3) -- (0.2, 4.3);
  \node[mutedgray, rotate=90, anchor=south] at (0.05, 2.3) {Qualitative Escalation};

  \draw[dashed, linkblue, thick]
    (1.1, 0.75) -- (3.0, 1.05) -- (5.0, 1.40) -- (7.0, 1.95) -- (9.1, 2.55) -- (11.1, 3.25) -- (12.8, 3.90);

  \fill[mutedgray] (1.1, 0.75) circle (3pt);
  \node[above=5pt] at (1.1, 0.75) {\texttt{210/248}};
  \node[below=5pt, align=center] at (1.1, 0.3) {\textbf{$mt{=}63$}\\Pfaffian};

  \fill[linkblue!70] (3.0, 1.05) circle (3pt);
  \node[above=5pt] at (3.0, 1.05) {\texttt{254/252/245}};
  \node[below=5pt, align=center] at (3.0, 0.3) {\textbf{$mt{=}96\text{--}110$}\\HOVER / L4};

  \fill[linkblue] (5.0, 1.40) circle (3.5pt);
  \node[pill, draw=linkblue, fill=boxbg, above=7pt] at (5.0, 1.40) {\texttt{TII-228} (18.9m)};
  \node[below=5pt, align=center] at (5.0, 0.3) {\textbf{$mt{=}126$}\\2.15 L40S-h};

  \fill[accentgreen] (7.0, 1.95) circle (4pt);
  \node[pill, draw=accentgreen, fill=greenbg, text=accentgreen, above=7pt] at (7.0, 1.95) {\texttt{TII-237} (Solved)};
  \node[below=5pt, align=center, text=accentgreen] at (7.0, 0.3) {\textbf{$mt{=}135$}\\22.1 L4-h};

  \fill[accentgreen] (9.1, 2.55) circle (4.5pt);
  \node[pill, draw=accentgreen, fill=greenbg, text=accentgreen, above=7pt] at (9.1, 2.55) {\texttt{TII-235} (Solved)};
  \node[below=5pt, align=center, text=accentgreen] at (9.1, 0.3) {\textbf{$mt{=}144$}\\39.6 L40S-h};

  \fill[accentgreen] (11.1, 3.25) circle (4.5pt);
  \node[pill, draw=accentgreen, fill=greenbg, text=accentgreen, above=7pt] at (11.1, 3.25) {\texttt{TII-185} (Solved)};
  \node[below=5pt, align=center, text=accentgreen] at (11.1, 0.3) {\textbf{$mt{=}153$}\\\texttt{v5} (140h+60h)};

  \fill[mutedgray!80!black] (12.8, 3.90) circle (3.8pt);
  \node[pill, draw=mutedgray, fill=boxbg, above=5pt] at (12.8, 3.90) {Classic McEliece};
  \node[below=5pt, align=center] at (12.8, 0.3) {\textbf{$mt{\ge}768$}\\348864+};
\end{tikzpicture}
\caption{Qualitative escalation of publicly documented TII Track-2 key recoveries by codimension $mt$. Our five verified recoveries (\texttt{TII-245}, \texttt{TII-228}, \texttt{TII-237}, \texttt{TII-235}, \texttt{TII-185}) span $mt=108 \to 153$, with \texttt{TII-185} extending past the $128$-bit lossless packing ceiling via truncated $16$-byte range pre-filtering (\texttt{v5}).}
\label{fig:ladder}
\end{figure}
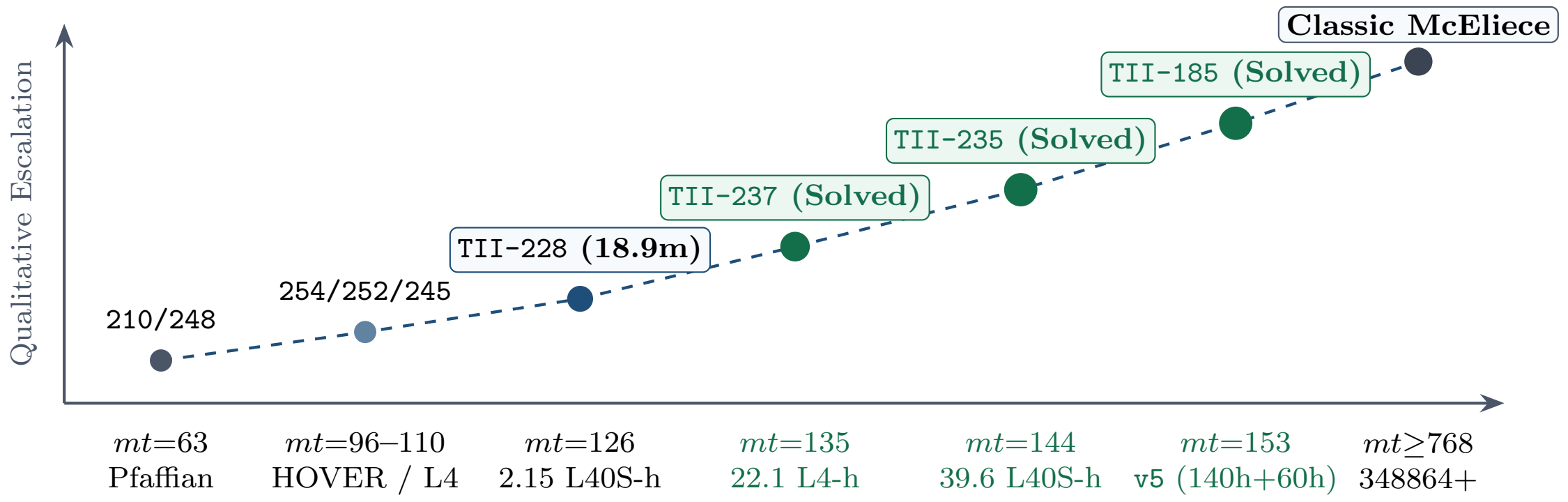

Primal-Shortened ISD~\cite{MiniMax26} inverts that trade-off. Rather than exploiting full support, it shortens the code further to $n'$ columns, harvests weight-$(2t+1)$ codewords, and recovers $(g, L)$ from quadratic polar forms. Puncturing no longer obstructs the attack, which explains why MiniMax's Route~A reached \texttt{TII-228} ($mt=126$, $12$ punctures, $2{,}463$ EPYC core-hours) and \texttt{TII-250} ($mt=120$, $66{,}722$ core-hours on $5{,}120\text{--}10{,}240$ cores). Figure~\ref{fig:ladder} and Table~\ref{tab:landscape} place our five solved instances along this escalation path, showing why TII challenge IDs (estimated under 2023 attacks) no longer match Primal-Shortened ISD cost.

\begin{table}[t]
\centering
\caption{Representative 2026 TII Track-2 key-recovery milestones and our five verified GPU recoveries.}
\label{tab:landscape}
\footnotesize
\setlength{\tabcolsep}{1.9pt}
\begin{tabular}{llllll}
\toprule
Tier & Instance & $(n, m, t)$ & $mt$ & Method & Compute / Status \\
\midrule
Prior & \texttt{TII-210/248} & $m=9, t=7$ & $63$ & Pfaffian / HOV~\cite{BLT26,Hemmert26} & CPU (minutes) \\
Prior & \texttt{TII-253} & $(214, 8, 9)$ & $72$ & Hold-out / GIJS~\cite{Vedenev26,Weis26} & CPU \\
Prior & \texttt{TII-254/252} & $m=8, 10$ & $96, 110$ & 2-Anchor / HOVER~\cite{Saarinen26HOVER,Saarinen26TwoAnchor} & $27.2$ GH200-h (\texttt{254}) \\
Control & \texttt{TII-245} & $(493, 9, 12)$ & $108$ & MiniMax / \textbf{Ours (1$\times$L4)} & $44.6$ Xeon-ch vs.\ \textbf{19.6\,m} \\
Benchmark & \texttt{TII-228} & $(500, 9, 14)$ & $126$ & MiniMax / \textbf{Ours (8$\times$L40S)} & $2{,}463$ EPYC-ch vs.\ \textbf{2.15 GPU-h} \\
New & \texttt{TII-237} & $(500, 9, 15)$ & $135$ & \textbf{Ours (Colab L4)} & \textbf{22.1 L4-h} + 23.2\,s lift \\
New & \texttt{TII-235} & $(502, 9, 16)$ & $144$ & \textbf{Ours (10$\times$L40S)} & \textbf{39.62 L40S-h} + 18.0\,s lift \\
New & \texttt{TII-185} & $(511, 9, 17)$ & $153$ & \textbf{Ours (\texttt{v5}: 7$\times$L40S+3$\times$L4)} & \textbf{140.4 L40S-h + 60 L4-h} \\
Margin & \texttt{348864+} & $(3488, 12, 64)$ & $\ge 768$ & Classic McEliece & Far beyond reach \\
\bottomrule
\end{tabular}
\end{table}

\section{Primal-Shortened ISD and Four GPU-Aware Modifications}

Let $L = (\alpha_0, \dots, \alpha_{n-1}) \in \mathbb{F}_{2^m}^n$ be an ordered support of $n$ distinct elements and $g(X) \in \mathbb{F}_{2^m}[X]$ a monic square-free Goppa polynomial of degree $t$ with $g(\alpha_i) \ne 0$ for all $i$. Taking a prefix $A = \{0, \dots, n'-1\}$ where $H_A \in \mathbb{F}_2^{mt \times n'}$ has full rank $mt$ (verified empirically on every instance) yields the shortened Goppa code $\mathcal{C}_A = \ker(H_A) = \Gamma(L_A, g)$ of dimension $k' = n' - mt$. Any codeword $c \in \mathcal{C}_A$ of minimum odd Hamming weight $w = 2t + 1$ and support $S \subset A$ has locator polynomial $\sigma_S(X) = \prod_{i \in S}(X - \alpha_i)$ satisfying the characteristic-$2$ derivative-square identity $\sigma'_S(X) = g(X)^2$ (attributed to Finiasz in~\cite{MiniMax26}). On the odd-weight affine slice of $\mathcal{C}_A$ (affine dimension $d = k' - 1$), every weight-$(2t+1)$ codeword satisfies $m+1$ quadratic relations over the $F(d) = 1 + d + \binom{d}{2}$ quadratic monomials in $d$ coordinates, including the public quadratic weight identity $\sum_{i<j} c_i c_j = \binom{w}{2} \bmod 2$ with associated symmetric bilinear polar form $T_{\mathrm{pub}}$~\cite{MiniMax26}. When harvested words span an evaluation matrix of rank $F(d) - (m+1)$, dense $\mathbb{F}_2$ elimination recovers the $(m+1)$-dimensional relation space. Modding out $T_{\mathrm{pub}}$ lifts the remaining $m$ polar forms to an equivalent $(L_A, g)$, and $O(n)$ field divisions then extend $L_A$ to all $n$ positions after applying the unscrambling transformation to $H_{\mathrm{pub}}$.

\begin{figure}[t]
\centering
\begin{tikzpicture}[
  >=Stealth,
  node font=\scriptsize,
  stage/.style={draw, rounded corners=3pt, minimum height=2.55cm, text width=3.20cm, align=center, inner sep=4pt}
]
  \node[stage, draw=accentgreen, fill=greenbg, thick] (s1) at (0,0)
    {\textbf{1. Optimal Shortening}\\[4pt]
     Widen $n'$ past CPU\\
     \texttt{228}: $395$, \texttt{235}: $425$\\
     $-34\%$ to $-47\%$ GPU-h\\[4pt]
     \textcolor{accentgreen}{\textbf{Ours (Mod.\ 4)}}};

  \node[stage, draw=accentgreen, fill=greenbg, thick] (s2) at (3.75,0)
    {\textbf{2. L2-Sized GPU-ISD}\\[4pt]
     16B zero-index packing\\
     $G \in \{4,8\}$ L2 tables\\
     Batched full-SM \texttt{k\_sys}\\[4pt]
     \textcolor{accentgreen}{\textbf{Ours (Mods.\ 1--3)}}};

  \node[stage, draw=linkblue, fill=boxbg] (s3) at (7.50,0)
    {\textbf{3. Polar Lifting}\\[4pt]
     $\mathbb{F}_2$ M4RI on $F(d)$ cols\\
     Quotient $m{+}1 \to m$\\
     $10\text{--}40\text{ s}$ in SageMath\\[4pt]
     \textcolor{linkblue}{MiniMax Route A~\cite{MiniMax26}}};

  \node[stage, draw=linkblue, fill=boxbg] (s4) at (11.25,0)
    {\textbf{4. Extend \& Verify}\\[4pt]
     $O(n)$ support extension\\
     Bit-exact RREF match\\
     \texttt{check\_solution} $= 1$\\[4pt]
     \textcolor{linkblue}{MiniMax + TII~\cite{TIIRepo}}};

  \draw[->, thick, mutedgray] (s1) -- (s2);
  \draw[->, thick, mutedgray] (s2) -- (s3);
  \draw[->, thick, mutedgray] (s3) -- (s4);
\end{tikzpicture}
\caption{End-to-end Primal-Shortened ISD pipeline. Shaded green boxes mark our four GPU-oriented modifications; blue boxes reuse MiniMax's algebraic lift and the unmodified official TII verification harness.}
\label{fig:pipeline}
\end{figure}
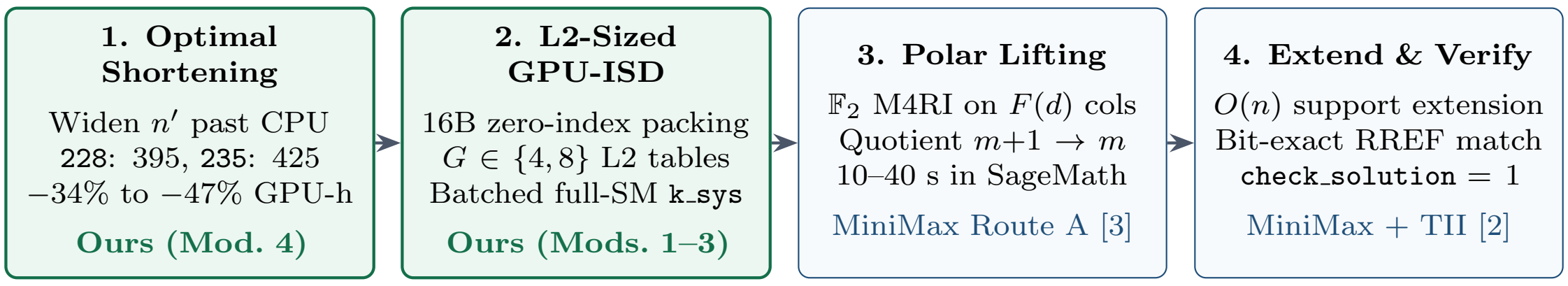

Figure~\ref{fig:pipeline} separates MiniMax's upstream algebraic pipeline from our four GPU-oriented modifications.

MiniMax's Mode-6 CPU sampler performs full random systematic reduction followed by a Stern-style~\cite{Stern89,Dumer91} meet-in-the-middle search with $p=3+3$ non-pivot bits, $l=17$ collision bits ($2^{17} = 131{,}072$ buckets), and a $16\text{ KB}$ occupancy bitmap pre-filter. On a CPU, each thread owns one reduction whose hash table fits in L2/L3 cache. Moving that layout directly to a GPU breaks down: our per-SM port (\texttt{v1}: $58$ blocks, $6$ slots $\times$ $32$ bytes per bucket, $25.7\text{ MB}$/table) allocates $1.49\text{ GB}$ of live hash tables. Shrinking entries to $16$ bytes with $4$ slots and raising occupancy to $2$ blocks/SM (\texttt{v2}: $116$ blocks, $1.04\text{ GB}$ total) lowered \texttt{TII-237} throughput from $2{,}208$ to $1{,}872\text{ it/s}$ because both allocations exceed the $50.3\text{ MB}$ L4 L2 cache by more than $20\times$ while twice as many blocks contend for cache lines. Our hardened solver \path{src/minwords_gpu_v4.cu} (derived from MiniMax's \texttt{minwords\_mt.c} under Apache-2.0) addresses this with four coupled modifications.

\paragraph{1. Lossless 16-byte packing without left indices.}
Because the bucket index $h \in [0, 2^{17})$ fixes the lowest $17$ syndrome bits, any instance with $mt \le 145$ packs all remaining $mt - 17 \le 128$ bits losslessly into two $64$-bit words:
\[
  p_0 = (s_0 \gg 17) \mid (s_1 \ll 47), \qquad p_1 = (s_1 \gg 17) \mid (s_2 \ll 47).
\]
We store only $(p_0, p_1)$ as a $16$-byte \texttt{ulonglong2} and drop left-triple indices $(a, b, c)$ from the hot bucket array. Under a uniform-syndrome heuristic given a $17$-bit bucket collision, the conditional weight-filter pass probability $\binom{mt-17}{w-6}/2^{mt-17}$ is $2^{-27.02}$ on \texttt{TII-245}, $2^{-31.39}$ on \texttt{TII-228}, $2^{-33.57}$ on \texttt{TII-237}, and $2^{-35.74}$ on \texttt{TII-235}. Because the $128$-bit packing is lossless up to $mt=145$, every exact match ($\operatorname{popc}(p_0 \oplus r_{p_0}) + \operatorname{popc}(p_1 \oplus r_{p_1}) = w - 6$) is a genuine weight-$w$ codeword of the systematic matrix; only on these rare hits does the thread unpack $(s_0, s_1, s_2)$ and recover $(a, b, c)$ from shared-memory \texttt{colmask}. With $4$ slots per bucket, each bucket payload is $64$ bytes aligned ($8.39\text{ MB}$/table, plus a separate $0.52\text{ MB}$ count array and $16\text{ KB}$ bitmap). Under a Poisson load model, the $4$-slot cap retains $98.03\%$ of left triples on \texttt{TII-245} ($\lambda \approx 1.60$), $95.13\%$ on \texttt{TII-237} ($\lambda \approx 2.20$), $89.4\%$ on \texttt{TII-228} at $n'=395$ ($\lambda \approx 2.99$), $85.8\%$ on \texttt{TII-235} at $n'=425$ ($\lambda \approx 3.41$), and $80.9\%$ on \texttt{TII-185} at $n'=447$ ($\lambda \approx 3.96$).

To cross the $mt=145$ packing ceiling without enlarging the $16$-byte bucket table, \path{src/minwords_gpu_v5.cu} raises $\texttt{MAXR}=192$ while continuing to pack bits $17\text{--}144$ into \texttt{ulonglong2}. For $145 < mt \le 192$, the $e = mt - 145$ omitted upper bits can contribute at most $e$ ones, so we apply the lossless range pre-filter $(w-6)-e \le \operatorname{popc}_{128} \le w-6$, reconstruct $(a, b, c)$ from the exact $145$-bit prefix, and evaluate the exact $mt$-bit weight against the full $3$-word \texttt{colmask}. On \texttt{TII-185} ($mt=153, e=8, w=35$), the range pre-filter passes with conditional probability $\sum_{j=21}^{29}\binom{128}{j}/2^{128} = 2^{-32.28}$ (vs.\ $2^{-37.91}$ for true hits), keeping false-positive reconstructions negligible while preserving the exact $35.7\text{ MB}$ / $71.4\text{ MB}$ L2-sized footprint.

\paragraph{2. Cooperative L2-sized execution ($G=4$ on L4, $G=8$ on L40S).}
Instead of assigning one table per SM, we allocate $G$ concurrent tables whose combined footprint stays below the target GPU's nominal L2 cache. On NVIDIA L4 ($58$ SMs, $50.3\text{ MB}$ L2) and A100 ($41.9\text{ MB}$ L2), $G=4$ with $16$ blocks per group ($64$ blocks total) uses $35{,}717{,}120$ bytes ($35.7\text{ MB}$). On NVIDIA L40S ($142$ SMs, $100.7\text{ MB}$ L2), compiling with \texttt{-DG\_GROUPS=8 -DBLOCKS\_PER\_GROUP=18} doubles concurrency to $G=8$ ($71.4\text{ MB} < 100.7\text{ MB}$ L2) and saturates all $142$ SMs with $144$ cooperative blocks. Every block caches its group's $16\text{ KB}$ occupancy bitmap in shared memory.

\paragraph{3. Batched systematic reduction and hardware scaling.}
Switching from \texttt{v2} to unbatched cooperative \texttt{v3} ($G=4$) lifted \texttt{TII-237} throughput to only $2{,}671\text{ it/s}$. CUDA event profiling over $256$ iterations on L4 explained the ceiling:
\[
  \texttt{k\_sys}: 68.00\text{ ms}~(74.0\%), \quad
  \texttt{k\_zero}: 0.34\text{ ms}, \quad
  \texttt{k\_build}: 10.82\text{ ms}, \quad
  \texttt{k\_probe}: 12.72\text{ ms}.
\]
Running \texttt{k\_sys<<<4, 256>>>} left $54$ of $58$ SMs idle during Gaussian elimination. Batching $64G$ independent reductions in one \texttt{k\_sys<<<64G, 256>>>} launch ($256$ reductions / $2.2\text{ MB}$ state buffer on L4; $512$ reductions / $4.4\text{ MB}$ on L40S) ahead of $64$ build/probe steps raised L4 throughput $3.5\times$ to $9{,}462\text{ it/s}$ on \texttt{TII-237} and $3.17\times$ to $11{,}035\text{ it/s}$ peak ($11{,}024\text{ it/s}$ full-run average) on \texttt{TII-245}. In our single-run \texttt{TII-237} comparison, an L4 ($50.3\text{ MB}$ L2, $2.04\text{ GHz}$ boost, $1.94\text{ CU/hr}$ on Colab) beat an A100-SXM4-40GB ($41.9\text{ MB}$ L2, $1.41\text{ GHz}$ boost, $\approx 6.8\text{ CU/hr}$, $8{,}035\text{ it/s}$) by $1.18\times$ in raw speed and $>4\times$ in iterations per CU. Scaling to L40S ($100.7\text{ MB}$ L2, $142$ SMs) reached $12{,}081\text{ it/s}$ on \texttt{TII-228} ($n'=395$) and $10{,}934\text{--}15{,}593\text{ it/s}$ (aggregate $11{,}684\text{ it/s}$) on \texttt{TII-235} ($n'=425$).

\paragraph{4. GPU-aware optimal shortening window $n'$.}
MiniMax's \path{optimal_shortening_sizes} picks a narrow prefix $n'$ whose random-code supply $N(n') = \binom{n'}{w}2^{-mt}$ exceeds the planning threshold $R(n') = F(k'-1)-(m+1)$ by roughly $1.4\text{--}2.2\times$, keeping the $\mathbb{F}_2$ linear system small. Under a Poisson-bucket and uniform coupon-collector model, expected iterations scale as
\[
  I(n') = \frac{-\ln\!\bigl(1 - 1.02\,R(n') / N(n')\bigr)}{\rho_4(\lambda)\,p_1(n')},
\]
where $\rho_4(\lambda)$ is the $4$-slot retention fraction at left-list load $\lambda = \binom{\lfloor k'/2\rfloor}{3}/2^{17}$ and $p_1(n') = \binom{\lfloor k'/2\rfloor}{3}\binom{\lceil k'/2\rceil}{3}\binom{mt-17}{w-6}/\binom{n'}{w}$. Modeled GPU-hours equal $I(n') \cdot \tau(n')$, with per-iteration time $\tau(n')$ scaled in proportion to the left-list size $\binom{\lfloor k'/2\rfloor}{3}$. A tight supply drives $-\ln(1 - R/N)$ sharply upward as duplicates accumulate. Widening $n'$ grows $R(n')$ only quadratically (SageMath M4RI still finishes in under $40\text{ s}$ across all five instances) while $N(n')$ grows rapidly as $\binom{n'}{w}$, cutting duplicate waste and raising per-iteration yield until $\binom{k'/2}{3}$ and $4$-slot overflow bend total GPU-hours back up (Figure~\ref{fig:bench}, right). On \texttt{TII-228}, widening $n'$ from a narrow $n'=355$ point to $n'=395$ lowers modeled L4 GPU-hours by $47\%$ ($7.4\text{ h} \to 3.9\text{ h}$; MiniMax's published \texttt{TII-228} run used $n'=361$). On \texttt{TII-235}, widening $n'$ from $393$ to $425$ lowers modeled L4 GPU-hours by $34\%$ ($122.0\text{ h} \to 80.5\text{ h}$), and in practice kept cross-worker overlap across $10$ parallel GPUs to $2.06\%$.

\section{Experiments}

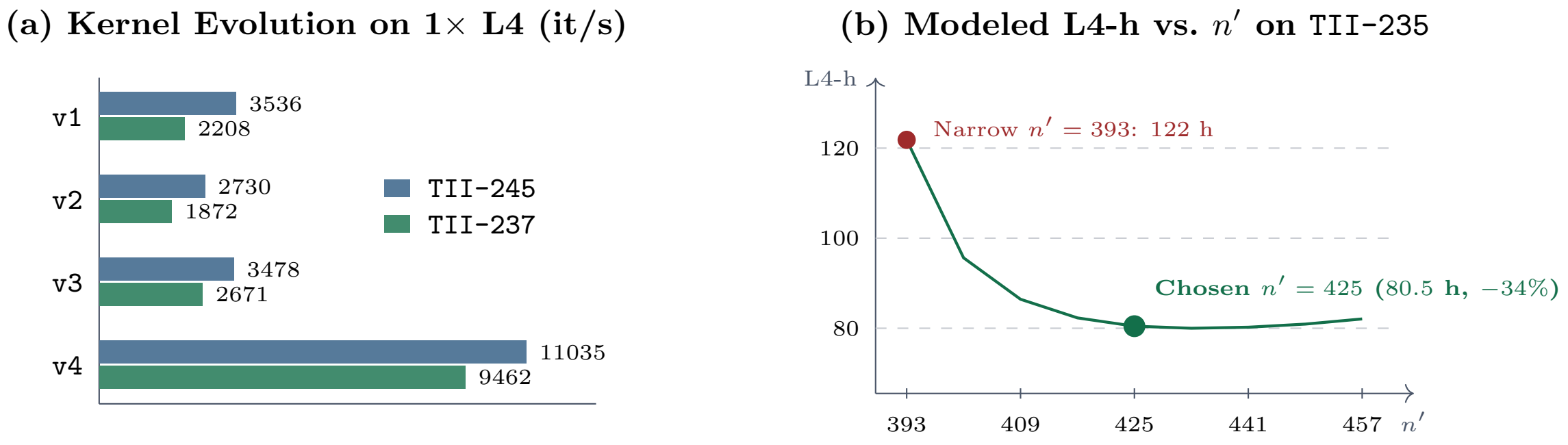
\begin{figure}[t]
\centering
\begin{tikzpicture}[node font=\scriptsize]
  \begin{scope}[xshift=0cm]
    \node[font=\footnotesize\bfseries] at (2.8, 3.95) {(a) Kernel Evolution on 1$\times$ L4 (it/s)};
    \draw[mutedgray] (0.7, 0.3) -- (0.7, 3.45);
    \draw[mutedgray] (0.7, 0.3) -- (5.5, 0.3);
    \foreach \v/\y/\ta/\tb in {v1/2.85/3536/2208, v2/2.05/2730/1872, v3/1.25/3478/2671, v4/0.45/11035/9462} {
      \node[anchor=east, font=\scriptsize\ttfamily] at (0.65, \y+0.22) {\v};
      \pgfmathsetmacro{\wa}{\ta/11500*4.3}
      \pgfmathsetmacro{\wb}{\tb/11500*4.3}
      \fill[linkblue!75] (0.7, \y+0.24) rectangle ({0.7+\wa}, \y+0.46);
      \node[anchor=west, font=\tiny] at ({0.72+\wa}, \y+0.35) {\ta};
      \fill[accentgreen!80] (0.7, \y) rectangle ({0.7+\wb}, \y+0.22);
      \node[anchor=west, font=\tiny] at ({0.72+\wb}, \y+0.11) {\tb};
    }
    \fill[linkblue!75] (3.45, 2.30) rectangle (3.70, 2.47);
    \node[anchor=west] at (3.77, 2.38) {\texttt{TII-245}};
    \fill[accentgreen!80] (3.45, 1.95) rectangle (3.70, 2.12);
    \node[anchor=west] at (3.77, 2.03) {\texttt{TII-237}};
  \end{scope}

  \begin{scope}[xshift=7.6cm]
    \node[font=\footnotesize\bfseries] at (3.1, 3.95) {(b) Modeled L4-h vs.\ $n'$ on \texttt{TII-235}};
    \draw[->, mutedgray] (0.6, 0.4) -- (5.8, 0.4) node[below=2pt, font=\tiny] {$n'$};
    \draw[->, mutedgray] (0.6, 0.4) -- (0.6, 3.45) node[left=2pt, font=\tiny] {L4-h};
    \foreach \h/\yy in {80/1.03, 100/1.90, 120/2.77} {
      \draw[mutedgray!30, dashed] (0.6, \yy) -- (5.6, \yy);
      \node[anchor=east, font=\tiny] at (0.55, \yy) {\h};
    }
    \foreach \np/\xx in {393/0.9, 409/2.0, 425/3.1, 441/4.2, 457/5.3} {
      \draw[mutedgray] (\xx, 0.35) -- (\xx, 0.45);
      \node[below=2pt, font=\tiny] at (\xx, 0.35) {\np};
    }
    \draw[thick, accentgreen]
      (0.90, 2.85) -- (1.45, 1.71) -- (2.00, 1.31) -- (2.55, 1.13) -- (3.10, 1.05) -- (3.65, 1.03) -- (4.20, 1.04) -- (4.75, 1.07) -- (5.30, 1.12);
    \fill[accentred] (0.90, 2.85) circle (2.5pt);
    \node[anchor=west, text=accentred, font=\tiny] at (1.05, 2.98) {Narrow $n'=393$: $122\text{ h}$};
    \fill[accentgreen] (3.10, 1.05) circle (3pt);
    \node[anchor=west, text=accentgreen, font=\tiny\bfseries, fill=white, inner sep=1pt] at (3.25, 1.42) {Chosen $n'=425$ ($80.5\text{ h}$, $-34\%$)};
  \end{scope}
\end{tikzpicture}
\caption{(a) Single-run \texttt{v1}--\texttt{v4} iteration throughput on one NVIDIA L4 (baseline sources in \texttt{src/baselines/}). (b) Modeled L4 GPU-hours vs.\ shortening prefix $n'$ for \texttt{TII-235} ($mt=144$), incorporating $4$-slot Poisson bucket retention and coupon-collector depletion.}
\label{fig:bench}
\end{figure}

Figure~\ref{fig:bench} summarizes both the \texttt{v1}--\texttt{v4} kernel progression on L4 and the modeled shortening-window trade-off for \texttt{TII-235}. All five CUDA kernels (\texttt{v1}--\texttt{v5}) are published in our repository.

\paragraph{Control run: \texttt{TII-245} ($n=493, m=9, t=12, mt=108$).}
At MiniMax's prefix $n'=326$, one NVIDIA L4 harvested $24{,}500$ host-verified weight-$25$ codewords in $1{,}175.5\text{ s}$ ($19.6\text{ min}$, $12{,}958{,}464$ iterations, $11{,}024\text{ it/s}$ average). SageMath 10.7 algebraic recovery on the Colab Xeon host completed in $9.78\text{ s}$ ($1{,}185.3\text{ s}$ harvesting plus lift), returning exact quotient dimension $10$, direct lift of dimension $9$, bit-exact RREF equality ($H_{\mathrm{cand}} = H_{\mathrm{pub}}$), and \texttt{check\_solution\_track2.sage} ($245$) $= 1$. MiniMax~\cite{MiniMax26} reported $2{,}500\text{ s}$ harvesting for $26{,}352$ words and $2{,}510\text{ s}$ end-to-end ($44.6$ core-hours) on a $64$-core Xeon Gold 6338 server ($2.12\times$ raw wall-clock ratio). Under uniform coupon-collector sampling with random-code supply $N = \binom{326}{25}/2^{108} \approx 52{,}299$, normalizing for their larger $26{,}352$-word target gives a $1.91\times$ speedup.

\paragraph{Head-to-head benchmark: \texttt{TII-228} ($n=500, m=9, t=14, mt=126$).}
MiniMax~\cite{MiniMax26} solved \texttt{TII-228} at $n'=361$ in $57.1\text{ min}$ wall-clock using $2{,}463$ physical CPU core-hours across $2{,}048\text{--}5{,}120$ AMD EPYC cores. Using our wider prefix $n'=395$ ($\operatorname{rank}(H_A)=126$, $k'=269$, threshold $36{,}037$), we ran eight Modal NVIDIA L40S workers ($G=8$, $144$ blocks, distinct seeds) from scratch, each targeting $4{,}720$ weight-$29$ codewords. Every worker finished with zero output-buffer overflows in $949.6\text{--}989.2\text{ s}$ ($12{,}025\text{--}12{,}125\text{ it/s}$, mean $12{,}081\text{ it/s}$), accumulating $93{,}574{,}144$ exact ISD iterations across $2.15$ L40S GPU-hours. End-to-end wall-clock time including Modal container startup and checkpoint transfer was $18.90\text{ min}$ ($\approx\$4.30$). Deduplicating the $37{,}760$ raw words yielded $37{,}084$ unique codewords ($1.79\%$ cross-worker overlap). SageMath 10.5 on an $8$-core CPU container completed the algebraic lift in $24.64\text{ s}$ ($19.94\text{ s}$ elimination, quotient dim $10$, direct lift dim $9$) and passed \texttt{check\_solution\_track2.sage} ($228$) $= 1$ in $5.43\text{ s}$. Although the two runs used different shortening prefixes and distinct hardware cost models (CPU core-hours vs.\ GPU-hours), eight L40S GPUs cut wall-clock recovery time on \texttt{TII-228} by $3.0\times$ relative to MiniMax's cluster run.

\paragraph{Codimension milestone: \texttt{TII-237} ($n=500, m=9, t=15, mt=135$).}
At $n'=378$ ($\operatorname{rank}(H_A)=135$, $k'=243$, threshold $29{,}394$, random-code supply $\approx 62{,}943$), we harvested $29{,}742$ unique weight-$31$ codewords across Colab L4 workers totaling $\approx 22.1$ GPU-hours and $\approx 7.4 \times 10^8$ iterations ($708.6\text{M}$ measured across five logged L4 workers plus $\approx 30\text{M}$ estimated on three short exploratory runs; full per-worker ledger in \path{codewords/RUN_MANIFEST.md}). Capture--recapture between our two largest independent-seed L4 workers ($|A|=17{,}009$, $|B|=8{,}570$, $|A \cap B|=2{,}215$) gave $\hat{N} = 65{,}809 \pm 1{,}123$ ($95\%$ CI $[63{,}608, 68{,}010]$), within $5\%$ of the random-code prediction. SageMath 10.7 lifted the key in $23.23\text{ s}$ ($15.33\text{ s}$ elimination, quotient dim $10$, direct lift dim $9$) and passed \texttt{check\_solution\_track2.sage} ($237$) $= 1$. Total Colab spend across all development, profiling, \texttt{TII-245}, and \texttt{TII-237} was $68.97\text{ CU}$.

\paragraph{Approaching the 128-bit packing ceiling: \texttt{TII-235} ($n=502, m=9, t=16, mt=144$).}
Punctured at $10$ positions ($512 - 502 = 10$) with Goppa degree $t=16$, \texttt{TII-235} sits well beyond current vanishing-attack parameters and requires $144 - 17 = 127$ packed syndrome bits---within one bit of our $128$-bit \texttt{ulonglong2} limit. At our widened prefix $n'=425$ ($\operatorname{rank}(H_A)=144$, $k'=281$, threshold $39{,}331$, random-code supply $\approx 787{,}006$), ten Modal NVIDIA L40S workers ($G=8$, $144$ blocks, distinct seeds) each targeted $4{,}120$ weight-$33$ codewords. All ten workers completed with zero overflows across $1.67 \times 10^9$ exact iterations and $39.62$ L40S GPU-hours in $4.30\text{ h}$ wall-clock ($\approx\$79$; Table~\ref{tab:results}). Merging the $41{,}200$ raw words produced $40{,}353$ unique codewords ($2.06\%$ overlap). Despite the widened $39{,}341$-feature matrix, SageMath 10.5 completed the algebraic lift in $17.97\text{ s}$ ($15.06\text{ s}$ elimination, exact quotient dim $10$, direct lift dim $9$), recovering a monic irreducible degree-$16$ polynomial $g(X) \in \mathbb{F}_{2^9}[X]$ and $502$-element support $L$ with bit-exact RREF match and \texttt{check\_solution\_track2.sage} ($235$) $= 1$ in $3.37\text{ s}$.

\paragraph{Crossing the 128-bit ceiling: \texttt{TII-185} ($n=511, m=9, t=17, mt=153$).}
At $mt=153$, \texttt{TII-185} exceeds our lossless $128$-bit packing format by $8$ bits and tests \path{src/minwords_gpu_v5.cu} directly. At widened prefix $n'=447$ ($\operatorname{rank}(H_A)=153$, $k'=294$, $F(293)=43{,}072$, threshold $43{,}062$, random-code supply $\approx 1{,}245{,}602$), we ran a hybrid $10$-GPU fleet combining seven Modal L40S workers ($G=8$, $144$ blocks, target $5{,}600$ each) and three Colab L4 workers ($G=4$, $64$ blocks). Across $21.1\text{ h}$ wall-clock on Modal ($140.4$ L40S GPU-hours, $\approx\$281$, $5.21 \times 10^9$ exact iterations, $35{,}726$ raw words) and $\approx 60.0$ L4 GPU-hours on Colab ($1.17 \times 10^9$ exact iterations, $8{,}031$ raw words), the fleet accumulated $6.38 \times 10^9$ exact ISD iterations and $43{,}757$ raw weight-$35$ codewords, deduplicating to $43{,}117$ unique codewords ($1.46\%$ cross-worker overlap; Table~\ref{tab:results}). SageMath 10.5 on an $8$-core CPU container completed the algebraic lift in $39.01\text{ s}$ ($31.60\text{ s}$ elimination, exact quotient dim $10$, direct lift dim $9$), recovering a monic irreducible degree-$17$ polynomial $g(X) \in \mathbb{F}_{2^9}[X]$ and $511$-element support $L$ with bit-exact RREF match and \texttt{check\_solution\_track2.sage} ($185$) $= 1$ in $6.12\text{ s}$. Table~\ref{tab:witnesses} records all fifteen SHA-256 verification digests.

\begin{table}[t]
\centering
\caption{Summary of all five verified GPU key recoveries. Total iterations for \texttt{TII-245}, \texttt{TII-228}, \texttt{TII-235}, and \texttt{TII-185} are exact device counters; \texttt{TII-237} combines $708.6\text{M}$ measured and $\approx 30\text{M}$ estimated iterations. For \texttt{TII-185}, wall-clock ($21.1\text{ h}$) reports the Modal fleet duration ($22.2\text{ h}$ including the Colab head start).}
\label{tab:results}
\small
\setlength{\tabcolsep}{3.5pt}
\begin{tabular}{lrrrrrrrr}
\toprule
Instance & $mt$ & $n'$ & Hardware & Wall & GPU-h & Iterations & Unique Words & Lift (s) \\
\midrule
\texttt{TII-245} & $108$ & $326$ & $1\times$ Colab L4 & $19.6\text{ m}$ & $0.33$ & $12.96\text{M}$ & $24{,}500$ & $9.78$ \\
\texttt{TII-228} & $126$ & $395$ & $8\times$ Modal L40S & $18.9\text{ m}$ & $2.15$ & $93.57\text{M}$ & $37{,}084$ & $24.64$ \\
\texttt{TII-237} & $135$ & $378$ & Colab L4 workers & --- & $\approx 22.1$ & $\approx 7.4\times 10^8$ & $29{,}742$ & $23.23$ \\
\texttt{TII-235} & $144$ & $425$ & $10\times$ Modal L40S & $4.30\text{ h}$ & $39.62$ & $1.67\times 10^9$ & $40{,}353$ & $17.97$ \\
\texttt{TII-185} & $153$ & $447$ & $7\times$L40S + $3\times$L4 & $21.1\text{ h}$ & $140.4{+}\!\approx\!60$ & $6.38\times 10^9$ & $43{,}117$ & $39.01$ \\
\bottomrule
\end{tabular}
\end{table}

\begin{table}[t]
\centering
\caption{Cryptographic verification digests ($\mathbb{F}_{2^9} = \mathbb{F}_2[a]/(a^9 + a^4 + 1)$). All five equivalent private keys pass the official TII SageMath checker; our pure-Python verifier additionally verifies all $174{,}796$ published codewords and SHA-256 digests.}
\label{tab:witnesses}
\scriptsize
\setlength{\tabcolsep}{4pt}
\begin{tabular}{lll}
\toprule
Instance & Artifact & SHA-256 Digest \\
\midrule
\texttt{TII-245} & \texttt{pk\_McEliece\_245.txt} & \texttt{ea029e86d34d13334bb9439f87333a8e87ebd902379ec515331d2bed78e973a8} \\
                 & \texttt{sk\_McEliece\_245.txt} & \texttt{967d2f4e934ca7617bdd71cfb5a0fcd98ec0f54a2c50db5170899ab3debac523} \\
                 & \texttt{words\_245\_np326.txt.gz} ($24{,}500$) & \texttt{b54b11b6c695b32a08aedb545cfb68dac9e0e3e7e1184175474344bfc59b4a11} \\
\midrule
\texttt{TII-228} & \texttt{pk\_McEliece\_228.txt} & \texttt{f98ddc98e6f54a60308b49ec36e16f70b6249ba4b7eac79f72cdf7fe70649f5a} \\
                 & \texttt{sk\_McEliece\_228.txt} & \texttt{717388cc390977ad536e23bdb6c7bc8e2cf69f065d69e3133f7fd27efd6dd51d} \\
                 & \texttt{words\_228\_np395.txt.gz} ($37{,}084$) & \texttt{46760cda2c768cdf3be5a559ed6a65895afe68d9e7241f52b91c80aede1e93d8} \\
\midrule
\texttt{TII-237} & \texttt{pk\_McEliece\_237.txt} & \texttt{b3cba630833e00d8e225d1943be87c31ddd74f270bd73f61c373aa70b1356f3b} \\
                 & \texttt{sk\_McEliece\_237.txt} & \texttt{128e26137b2b2e8128316cbc5e1c7766073d790e26a308a9766a120f05770e0f} \\
                 & \texttt{words\_237\_np378.txt.gz} ($29{,}742$) & \texttt{7793b74438d76235c8f5048a6b5644113b8372d999f837e1c09c274b2a472531} \\
\midrule
\texttt{TII-235} & \texttt{pk\_McEliece\_235.txt} & \texttt{f56256c7b6f253ff7748207ee05545d92644f6f8c5a7189fcaff9442eedd87a5} \\
                 & \texttt{sk\_McEliece\_235.txt} & \texttt{e96fdd90a9364e275d110dfb20b3fc56553123a7fb9d65e4c64772dbc195c221} \\
                 & \texttt{words\_235\_np425.txt.gz} ($40{,}353$) & \texttt{e8ee29c0b4750d3da4ac48b9e27d99c9c14598dad46bcf6526cd093e35d5329c} \\
\midrule
\texttt{TII-185} & \texttt{pk\_McEliece\_185.txt} & \texttt{19d1b2d4425b0541bb2f9061f3d3271c69159b95ce961014fefd79499293b140} \\
                 & \texttt{sk\_McEliece\_185.txt} & \texttt{adf36e4b00a35549e887a25793e2361b0ea53dbbae1ec888b7317a8532f1a9db} \\
                 & \texttt{words\_185\_np447.txt.gz} ($43{,}117$) & \texttt{5c9a58e7361707792d2b0e72c807238176049e545524c9c0d6b38af4e5670e55} \\
\bottomrule
\end{tabular}
\end{table}

\section{Limitations and Artifacts}
\label{sec:limits}

\paragraph{Cache residency and hardware variance.}
Sizing the table footprint below nominal L2 capacity ($35.7\text{ MB}$ on L4, $71.4\text{ MB}$ on L40S) does not prove $100\%$ L2 hit rates without Nsight hardware counters, and the \texttt{v1}--\texttt{v3} and A100 comparisons are single-run observations without $G \in \{2, 4, 8\}$ ablations on identical hardware. On Modal, L40S nodes exhibited two distinct boost-clock tiers ($11.0\text{k}$ vs.\ $15.6\text{k it/s}$ on \texttt{TII-235}), which accounts for the $3.0\text{--}4.3\text{ h}$ spread across workers.

\paragraph{Sampling bias and non-determinism.}
Non-pivot columns are partitioned in sorted order rather than shuffled per reduction, introducing sampling bias; $4$-slot buckets discard left triples in atomic arrival order ($\approx 4.9\%$ on \texttt{TII-237}, $\approx 14.2\%$ on \texttt{TII-235}, and $\approx 19.1\%$ on \texttt{TII-185}), making per-iteration output non-deterministic. Both effects render capture--recapture supply estimates heuristic.

\paragraph{Domain bounds and key equivalence.}
\path{src/minwords_gpu_v4.cu} enforces $mt \le 145$ (lossless) and \path{src/minwords_gpu_v5.cu} extends to $mt \le 192$ via truncated $16$-byte range pre-filtering, both with $k' \le 320$ and $\texttt{mode}=6$. Moving past $mt=153$ hits either $m=9, t=18$ ($mt=162$) or $m=10, t=16$ ($mt=160$, $k' \approx 397$), where modeled ISD search cost climbs by $5\times$ to $>100\times$. All recovered witnesses certify \emph{equivalent} private keys $(g, L)$ generating $H_{\mathrm{pub}}$ bit-for-bit, not necessarily the originally sampled generator representation.

\paragraph{Artifacts.}
All CUDA sources (\texttt{v1}--\texttt{v5}), compressed codeword datasets ($174{,}796$ verified words across five instances), equivalent private keys, and run manifests are available at \url{https://github.com/raullenchai/tii-mceliece-gpu} under Apache-2.0 (preserving MiniMax's notices in \path{LICENSE} and \path{NOTICE}). Running
\begin{center}
\texttt{python3 src/verify\_witnesses.py -{}-check-words}
\end{center}
checks all five equivalent private keys, hardcoded SHA-256 digests, and all $174{,}796$ harvested codewords in pure Python with zero external dependencies, while \path{src/solve_from_words.py} reproduces the SageMath algebraic recovery directly from the published codeword archives.

\section{Conclusion}

Packing MiniMax's Mode-6 Stern harvester into the GPU L2 cache, widening the shortening window $n'$, and extending past $mt=145$ via truncated $16$-byte range pre-filtering moves high-codimension Primal-Shortened ISD from multi-thousand-core CPU clusters to a handful of serverless GPUs. Eight Modal L40S GPUs solve \texttt{TII-228} ($mt=126$) in $18.9\text{ min}$ ($2.15$ GPU-hours), Colab L4 workers solve \texttt{TII-237} ($mt=135$) in $\approx 22.1$ GPU-hours, ten Modal L40S GPUs recover \texttt{TII-235} ($mt=144$) in $4.30\text{ h}$ ($39.62$ GPU-hours, $\approx\$79$), and a hybrid $7\times$ L40S + $3\times$ L4 fleet recovers \texttt{TII-185} ($mt=153$) in $21.1\text{ h}$ Modal wall-clock ($140.4$ L40S + $\approx 60$ L4 GPU-hours, $39.01\text{ s}$ lift). These experiments do not demonstrate practical key recovery at proposed Classic McEliece parameter sets ($m \ge 12, t \ge 64, mt \ge 768$).

\section*{Acknowledgments}
We thank Modal (\url{https://modal.com}) for providing serverless NVIDIA L40S GPU infrastructure for our \texttt{TII-228}, \texttt{TII-235}, and \texttt{TII-185} experiments, Google for providing NVIDIA L4 and A100 GPU compute through Google Colab for our \texttt{TII-245}, \texttt{TII-237}, \texttt{TII-185}, and kernel-profiling experiments, and MiniMax for releasing their Primal-Shortened ISD framework and Mode-6 sampler under Apache-2.0.

\end{document}